\documentclass{PoS}
\usepackage{multirow}
\usepackage{booktabs}

\newcommand{\mev}{\mathrm{MeV}}
\newcommand{\gev}{\mathrm{GeV}}
\newcommand{\tev}{\mathrm{TeV}}

\newcommand{\cm}{\mathrm{cm}}

\newcommand{\mum}{\mathrm{\mu m}}

\newcommand{\pt}{p_{\rm t}}

\newcommand{\lsim}{~{\buildrel < \over {_\sim}}~}
\newcommand{\gsim}{~{\buildrel > \over {_\sim}}~}
\newcommand{\dokpi}{{\rm D}^{0}\rightarrow {\rm K}^{-}\pi^{+}}
\newcommand{\dpluskpipi}{{\rm D}^{+}\rightarrow {\rm K}^{-}\pi^{+}\pi^{-}}
\newcommand{\dstardopi}{{\rm D}^{\star+}\rightarrow {\rm D}^{0}\pi^{+}_{\rm soft}}
\newcommand{\dokpipipi}{{\rm D}^{0}\rightarrow {\rm K}^{-}\pi^{+}\pi^{-}\pi^{+}}

\newcommand{\pp}{pp~}

\newcommand{\JINST}{\emph{JINST}~}
\newcommand{\EPJC}{\emph{Eur.~Phys.~J.}~C~}

\newcommand{\dorphi}{d_{0}(\mathrm {r}\phi)}
\newcommand{\evmult}{N_{\rm ch}/{\rm ev}}
\newcommand{\dedx}{{\rm d}E/{\rm d}x}
\newcommand{\dxy}{\Delta xy\vert_{y=0}} 
\newcommand{\dxloc}{\Delta x_{\rm loc}} 

\title{Alice Alignment, Tracking and Physics Performance Results}

\ShortTitle{Alice Alignment, Tracking and Physics Performance Results}

\author{\speaker{Andrea Rossi}\\
	University of Padova and INFN\\
	E-mail: \email{rossia@pd.infn.it}}
\author{for the ALICE Collaboration}

\abstract{The ALICE detector was designed to track and identify
  particles in a wide transverse momentum range, from more than $100~\gev/c$
  down to $\sim 100~\mev/c$. 
  The innermost barrel-like detector, the Inner Tracking System (ITS), 
  is dedicated to precise tracking
  and to primary and secondary vertices recontruction. 
  In this proceeding we present
  the ITS performance for tracking, primary vertex reconstruction
  and particle identification capability. 
  The current understanding of the detector response 
  is the result of an extensive phase of commissioning
  started with 2008 cosmic-ray run and finalized
  with data from \pp collisions: the alignment status of the
  three ITS sub-systems (Silicon Pixel Detectors, Silicon Drift Detectors, Silicon Strip Detectors) and 
  the study of the detector material budget via the
  reconstruction of gamma conversion in the material are presented.
  Precise tracking is required 
  to reconstruct and separate the primary vertex of interaction
  from secondary vertices from hyperons and heavy-flavour meson decays: 
  some examples of the achieved results are shown.
}

\FullConference{19th International Workshop on Vertex Detectors\\
          June 6 -11 2010\\
          Loch Lomond, Scotland, UK}

\begin{document}

\section{Introduction}
The ALICE detector is composed of two main sections: a central barrel, covering the full azimuth
in the acceptance region $|\eta|<0.9$ and a forward ($2.5<\eta<4$)
muon arm ~\cite{aliceJINST}. 
The ITS (Inner Tracking System) is a cylindrically-shaped silicon 
tracker in the central barrel that surrounds the interaction region.
Its main task is to provide precise track and vertex reconstruction
close to the interaction point.
In particular, 
the ITS was designed with the aim to improve the position, angle, and momentum resolution for 
tracks reconstructed in the Time Projection Chamber (TPC, inner radius~80~$\cm$), 
to identify the secondary vertices from the decay of
hyperons and heavy flavoured hadrons,
to reconstruct the interaction vertex 
with a resolution better
than 100~$\mu$m, and 
to recover 
particles that are missed by the TPC due to acceptance limitations. 

In this proceeding we present
the ITS performance for tracking, primary vertex reconstruction
and particle identification capability. 
We start by describing the ITS layout in Section~\ref{sec:detector_ITSdescription}.
To preserve the design performance
and to tune MC simulations in order to reproduce as realistically as possible the detector 
response and geometry, alignment and a good knowledge of the material budget are required.
These are addressed in Section~\ref{ITStrackvertexPerformance}. The actual 
tracking and vertexing performance is reported in the same section: the resolution
on the impact parameter in the transverse plane is used as a benchmark quantity
to summarize the results. 
The ITS capability for particle identification at low momentum is
described in Section~\ref{sec:ITSpid}.
In Section~\ref{sec:physperformance} results
of physics analyses based on the precise track and vertex reconstruction
guaranteed by the ITS are shown. 
\begin{figure}[!t]
\centering
\vspace{-6mm}
\resizebox{0.5\textwidth}{!}{%
  \includegraphics*[]{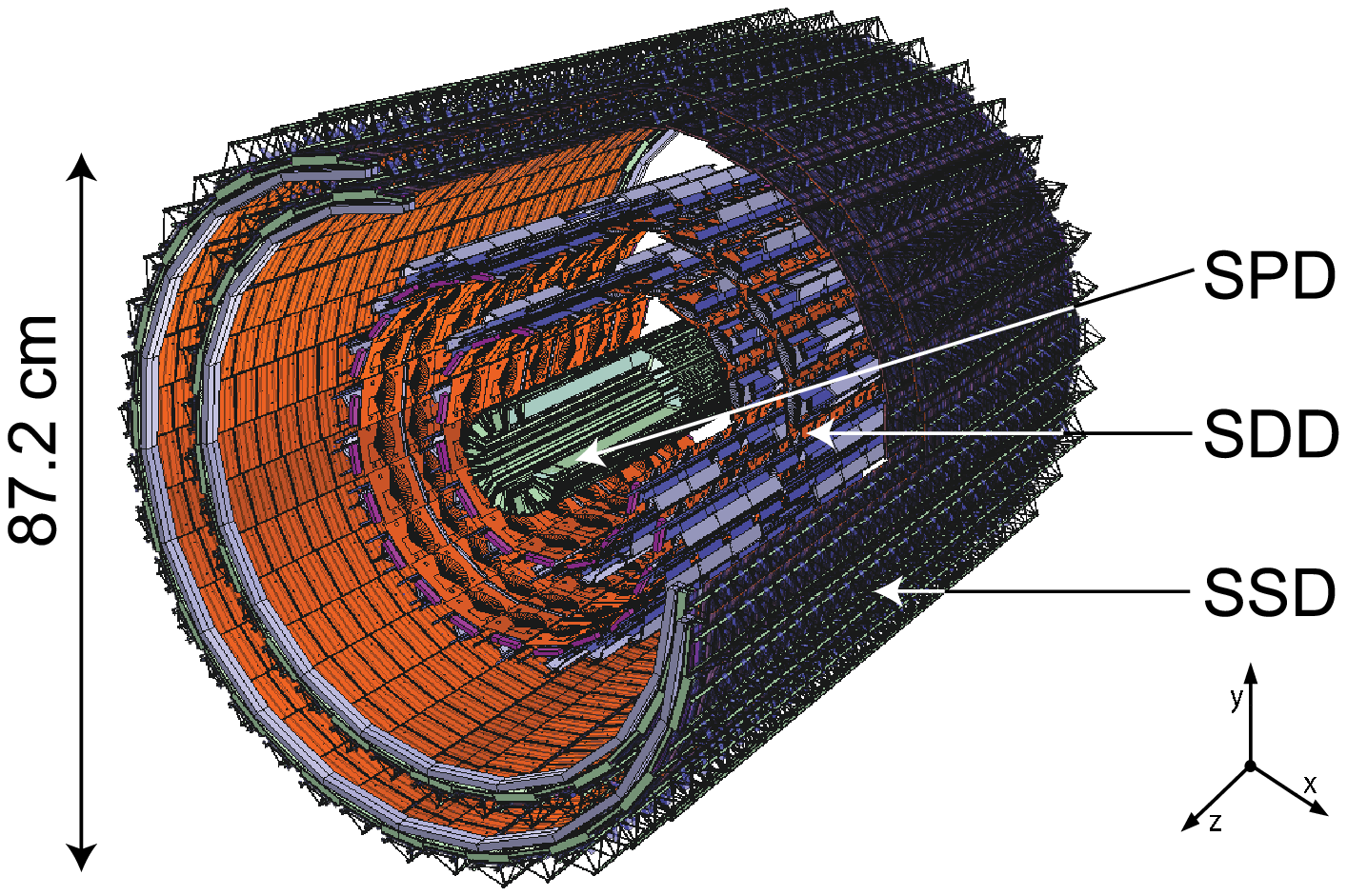}}
\resizebox{0.40\textwidth}{!}{%
  \includegraphics[]{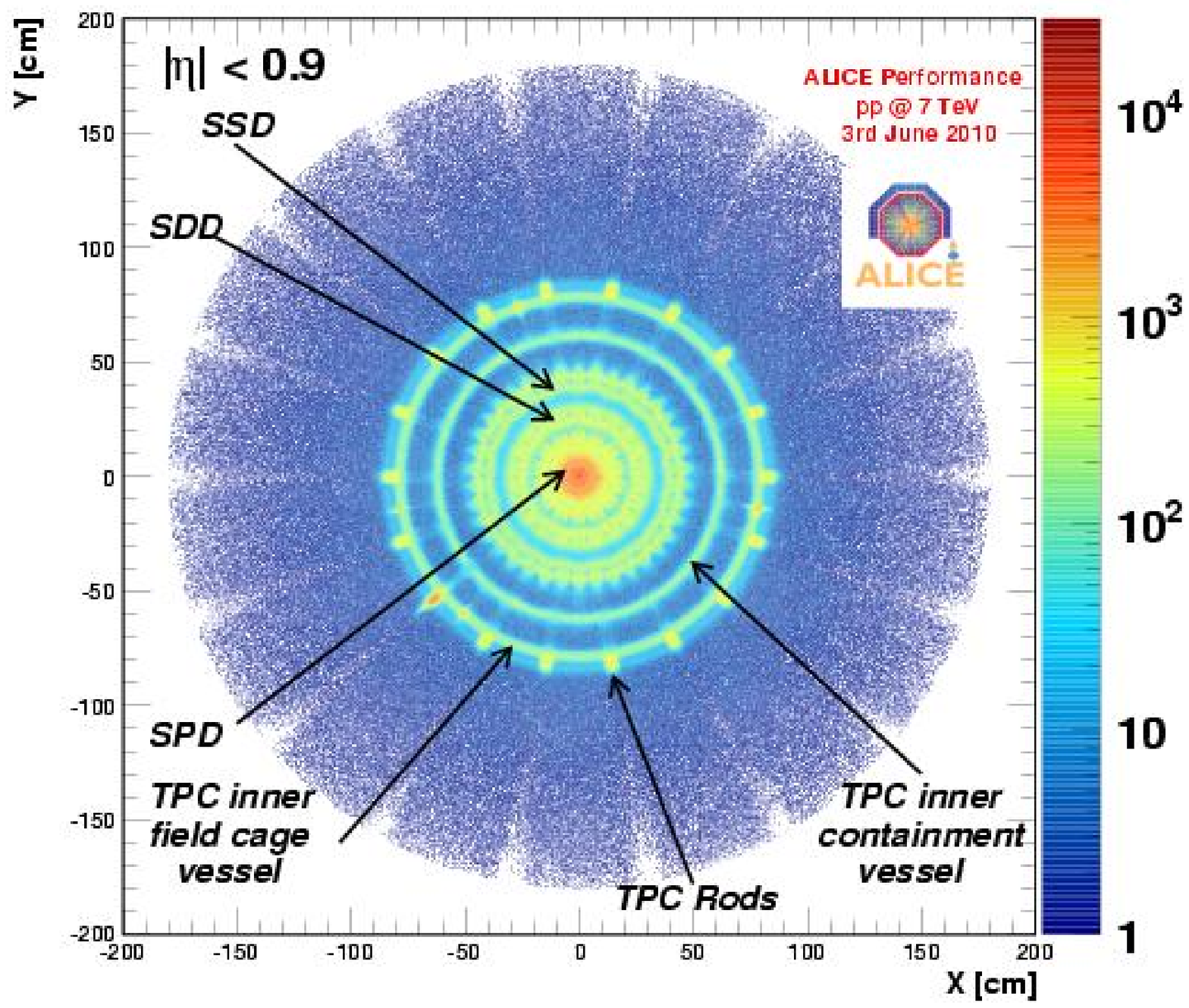}
}
\caption{(left) layout of the ITS and definition of the ALICE global 
  reference system. Right: two-dimensional ({\it xy}) distribution of gamma conversion points obtained
  with data from \pp collisions at 7~$\tev$. 
}
\label{fig:detector_itsscheme_Conversion}
\end{figure}
\section{Layout of the ALICE Inner Tracking System}
\label{sec:detector_ITSdescription}
The ITS consists of six layers,
with radii between 3.9~cm and 43.0~cm, covering
a pseudorapidity range varying from $|\eta|<0.98$ (outermost layer)
up to $|\eta|<1.98$ (innermost layer). 
The two innermost layers are equipped with Silicon Pixel Detectors (SPD), 
the two intermediate layers contain Silicon Drift Detectors (SDD), while
Silicon Strip Detectors (SSD) are used on the two outermost layers.
The geometrical layout of the ITS layers, as implemented in the 
ALICE simulation and reconstruction software framework (AliRoot~\cite{aliroot}),
is shown in Fig.~\ref{fig:detector_itsscheme_Conversion} (left).
The module local reference system 
is defined with the $x_{\rm loc}$ and $z_{\rm loc}$ axes on 
the sensor plane and the $z_{\rm loc}$ axis in the same direction as the
global $z$ axis. The local $x$ direction is approximately equivalent to 
the global $r\varphi$ (${\rm d}x_{\rm loc}\approx {\rm rd}\varphi$).
The geometrical parameters of the layers 
are summarised in Table~\ref{tab:ITSlayers}.
As far as the material budget is concerned, it should be noted that the values
reported in Table~\ref{tab:ITSlayers} account for sensor, electronics,
cabling, support structure and cooling for particles crossing the ITS
perpendicularly to the detector surfaces. 
Another 1.30\% of radiation length comes from the thermal
shields and supports installed between SPD and SDD barrels and between SDD
and SSD barrels, thus making the total material budget for perpendicular 
tracks equal to 7.66\% of $X_0$.
\begin{table}[t!]
  \caption{Characteristics of the six ITS layers.} 
  \label{tab:ITSlayers}
  \centering
  \begin{tabular}{|c|l|c|c|c|c|c|c|} 
    \hline 
    & & & & Number &
    Active Area & & Material\\
    Layer  & Type     &  $r$ [cm]  & $\pm z$ [cm]              & of   &
    per module  & Resolution & budget\\
    &      &          &              &    modules &
    $r\varphi$~$\times$~$z$ [mm$^2$] & $r\varphi$~$\times$~$z$ [$\mum^2$] & $X/X_0$ [\%] \\
    \hline
    \hline
    1-2  & pixel & \phantom{0}3.9-7.6 & 14.1  & \phantom{0}80-160 & 
    12.8$\times$70.7 & \phantom{0}12$\times$100 & 1.14\\
    3-4  & drift & 15.0-23.9           & 22.2-29.7  & \phantom{0}84-176  &
    70.17$\times$75.26 & \phantom{0}35$\times$25\phantom{0} & 1.13-1.26\\
    5-6  & strip & 38.0-43.0         & 43.1-48.9  & 748-950   &
    73$\times$40 & \phantom{0}20$\times$830        & 0.83-0.86 \\
    \hline
  \end{tabular}
\end{table}
For a detailed description of the features of the
three sub-detectors (SPD, SDD and SSD) and their status
see~\cite{aliceJINST,TurrisiThis,SittaThis}.
\section{ITS performance for track and vertex reconstruction}
\label{ITStrackvertexPerformance}
The precise reconstruction of track trajectory, of beam interaction point
and of decay points of unstable particles 
is a key element for ALICE, of fundamental importance for the fulfilment of its physics programme. 
A benchmark quantity to represent track and vertex reconstruction performance is
the resolution on the impact parameter in the transverse plane, $\dorphi$,
defined as the distance between the projection of a track in the transverse plane
and the reconstructed position of the primary vertex. 
For instance, most of heavy-flavour analyses rely on the capability 
of identifying secondary tracks, coming from charm and beauty hadron decays, displaced from the primary vertex. 
The ``resolving power'' needed is dictated by the impact parameter typical
of these tracks: as an example, $\langle \dorphi \rangle \approx {\rm c}\tau (D^{0})\approx 124~\mum$ 
for the kaon and pion tracks coming from a $\dokpi$ decay. 
This value must be compared with the impact parameter resolution, 
which is determined by the precision of track reconstruction
and extrapolation to the primary vertex. 
Due to multiple scattering, the material budget
determines the transverse momentum trend of the impact parameter resolution
and the dependence on the particle mass. 
The detector resolution, along with the detector layout,  
adds a $\pt$ independent contribution determining the asymptotic value of the resolution. 
The intrinsic detector resolution, determined by the sensor
structure and response, accounts for the uncertainty on the hit position
on the sensor. A further contribution derives from misalignment, that is, from
the uncertainty of the detector position in the global reference 
system where tracking is performed. 
The track model adopted for track reconstruction, in particular relative to the
treatment of the interaction with the material (energy loss, multiple scattering) 
can influence the reconstruction and extrapolation of track trajectory. 
The contribution of the uncertainty on the reconstructed primary vertex 
position depends on the event multiplicity ($\evmult$)
and is generally not constant in $\pt$ (high $\pt$ tracks are produced 
more copiously in high multiplicity events).
MC simulations must reproduce as realistically as possible the detector 
response and geometry, the material budget and the residual misalignment to 
prevent large systematic errors to arise. In this section all the above issues are addressed.
\subsection{Study of material budget amount with gamma conversion}
The dependence of photon
conversion probability on the radiation thickness of a detector element
has been exploited to obtain a detailed ``tomography'' of the detector 
by studying the distribution of the conversion points. 
In Fig.~\ref{fig:detector_itsscheme_Conversion} (right) the two-dimensional ({\it xy}) 
distribution of gamma conversion points obtained with \pp collisions
at $7~\tev$ is shown. Gamma conversion candidates are obtained via the V0 topology identification 
which is performed during the track reconstruction. 
The electron and positron daughter tracks are further selected using the TPC PID information. 
Vertex and mass constraints 
are applied as well. The coordinates of the conversion point are 
recalculated imposing that the daughter tracks are parallel at the conversion point. 
The current simulation reproduces the amount and the spatial distribution of reconstructed 
conversion points in great detail, with a relative accuracy of
a few percent. 
\subsection{Alignment of the Inner Tracking System}
%
%
The position and orientation in space 
of each of the 2198 ITS modules are defined by the six parameters 
of a roto-translation: more than 13000 parameters
must be determined to align the entire ITS. 
The sources of alignment information 
are the survey measurements (for SDD and SSD) and the reconstructed space points from
cosmic-rays and collision particles. These points
are the input for the software alignment methods, based on global or
local minimization of the residuals. 

The adopted strategy for the ITS is outlined as follows:
the first step is the validation of the SSD survey measurements with cosmic-ray tracks;
then the alignment of SPD and SSD is performed with cosmic-ray tracks, without magnetic field;
the already aligned SPD and SSD are used to confirm and refine 
the initial time-zero calibration of SDD, obtained with SDD standalone 
methods; in the meanwhile the SDD survey measurements with cosmic-ray tracks are validated;
then the full detector (SPD, SDD, SSD) is aligned with cosmic-ray tracks,
including also data collected with magnetic field $\rm B=0.5~T$;
tracks from pp collisions with both $\rm B=0$ and $\rm B=0.5~T$
are used together with tracks from cosmic-rays to complete and improve the alignment.
The relative alignment of the ITS to the TPC is performed in the meanwhile. 

The results achieved with cosmic-ray data collected
in 2008 
are described in~\cite{ITSalignArticle,noteITSalign,SSDnote}. About $10^{5}$ tracks 
from cosmic-ray events with magnetic field off were considered for the ITS alignment.
Cosmic-ray tracks have a dominant vertical component and the 
sides of the barrel layers have limited statistics. 
Tracks from proton-proton collisions, with both magnetic field off and $\rm B=0.5~T$, 
are now used to align the remaining
SPD modules (poorly illuminated by cosmic-rays) and to monitor and improve the alignment
quality of the whole ITS.

The following three observables are used to check the quality of the alignment:
the track-to-point residual, the track-to-point distance for ``extra'' points
in the acceptance overlaps and, with cosmic-ray tracks only, 
the top half-track to bottom half-track matching at the plane $y=0$ ($\equiv \dxy$). 
The latter provides a direct measurement of the resolution on
the track transverse impact parameter $\dorphi$, 
namely $\sigma_{\dxy}(\pt)=\sqrt{2}\sigma_{\dorphi}(\pt)$.
The pairs of points produced by particles crossing the acceptance overlap between 
two neighbouring modules allow us to
verify the relative position of the modules themselves. On average,
they allow to estimate the effective
spatial resolution of the sensor modules, 
i.e. the combination of the intrinsic spatial resolution and the
residual misalignment. 
Since the two ``extra'' points
are rather close in space and the amount of material crossed by the particle 
in-between the two points is very limited, multiple scattering can be neglected for tracks
of high enough momentum. 
The dependence of the 
intrinsic sensor resolution on the track-to-module incidence angle 
has to be accounted for. A different incidence angle implies a different
path in the silicon material and, depending also on electronic thresholds,
a different cluster shape.
The error on $\dxloc$ can be related to the effective spatial resolution 
of the two points, $\sigma_{\rm spatial}$, as:
\begin{equation}
  \sigma_{\dxloc}^2=\sigma^{2\,\rm eff}_{{\rm spatial} x}(\alpha_2)+ \sigma^{2\,\rm eff}_{{\rm spatial} x}(\alpha_1)\cos^2(\varphi_{12})
  \label{sigma_ovl}
\end{equation}
where the 1 and 2 subscripts indicate the two overlapping points,
$\alpha_i$ is the (unsigned) incidence angle of the track on the module plane 
and $\varphi_{12}$ is the relative angle between the two module planes. For overlaps
between modules on the same ladder (i.e. along the {\it z} direction) 
$\varphi_{12}=0^{\circ}$ while, for overlaps between modules on different ladders
(i.e. in $r\varphi$), $\varphi_{12} \approx 18^{\circ}$ and $9^{\circ}$ 
in the inner and outer SPD, $25.7^{\circ}$ and $16.4^{\circ}$ in the inner and outer
SDD, $10.6^{\circ}$ and $9.5^{\circ}$ in the inner and outer SSD.

In the following we briefly review the alignment status of the three
sub-detectors. The average residual misalignment is summarized by the values reported
in the second row of Table~\ref{tab:ResolMisalParamTracking}. 
\begin{figure}[!t]
  \begin{center}
    \vspace{-4mm}
      \includegraphics*[height=.24\textheight]{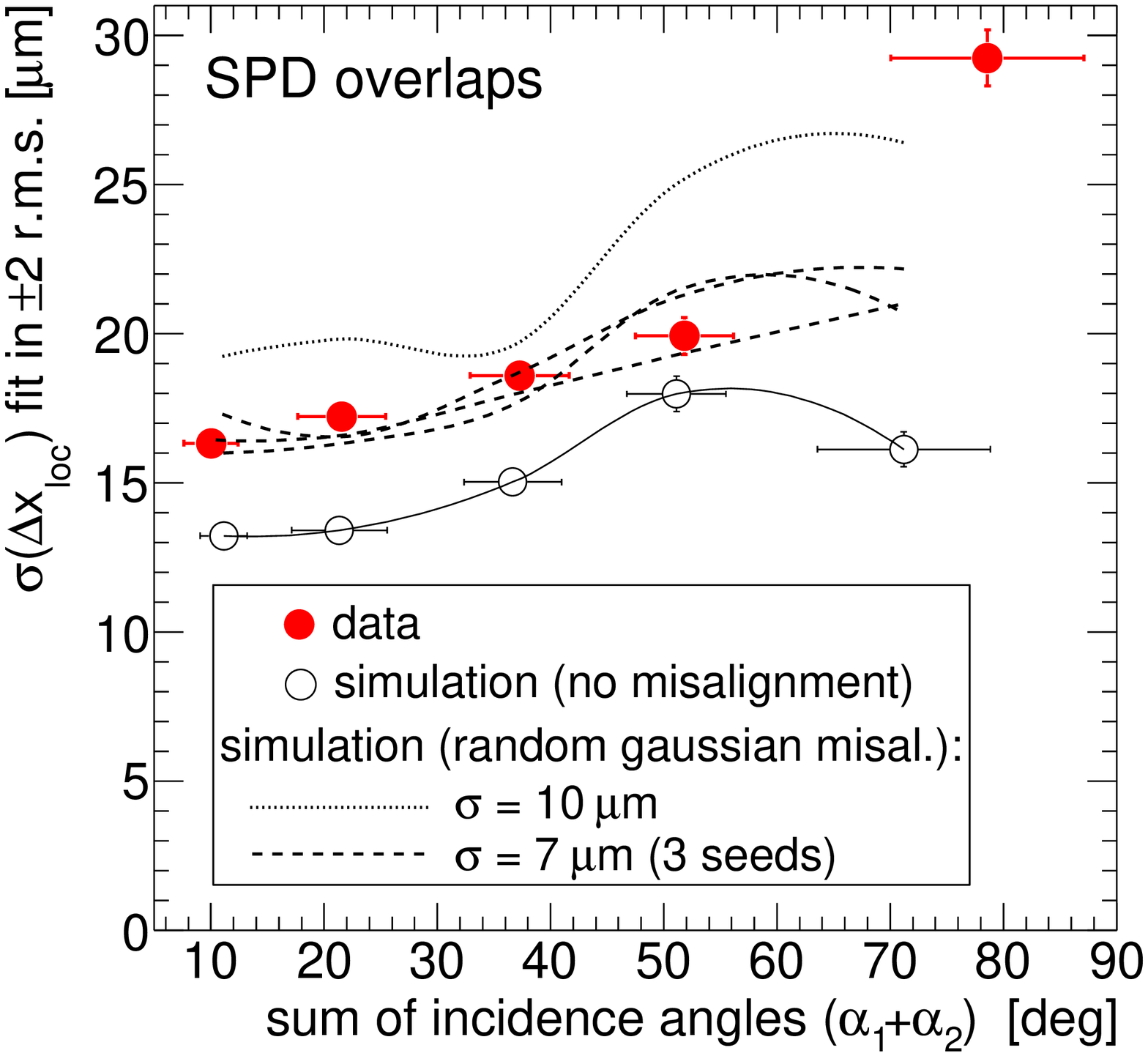}   
      \includegraphics[height=.26\textheight]{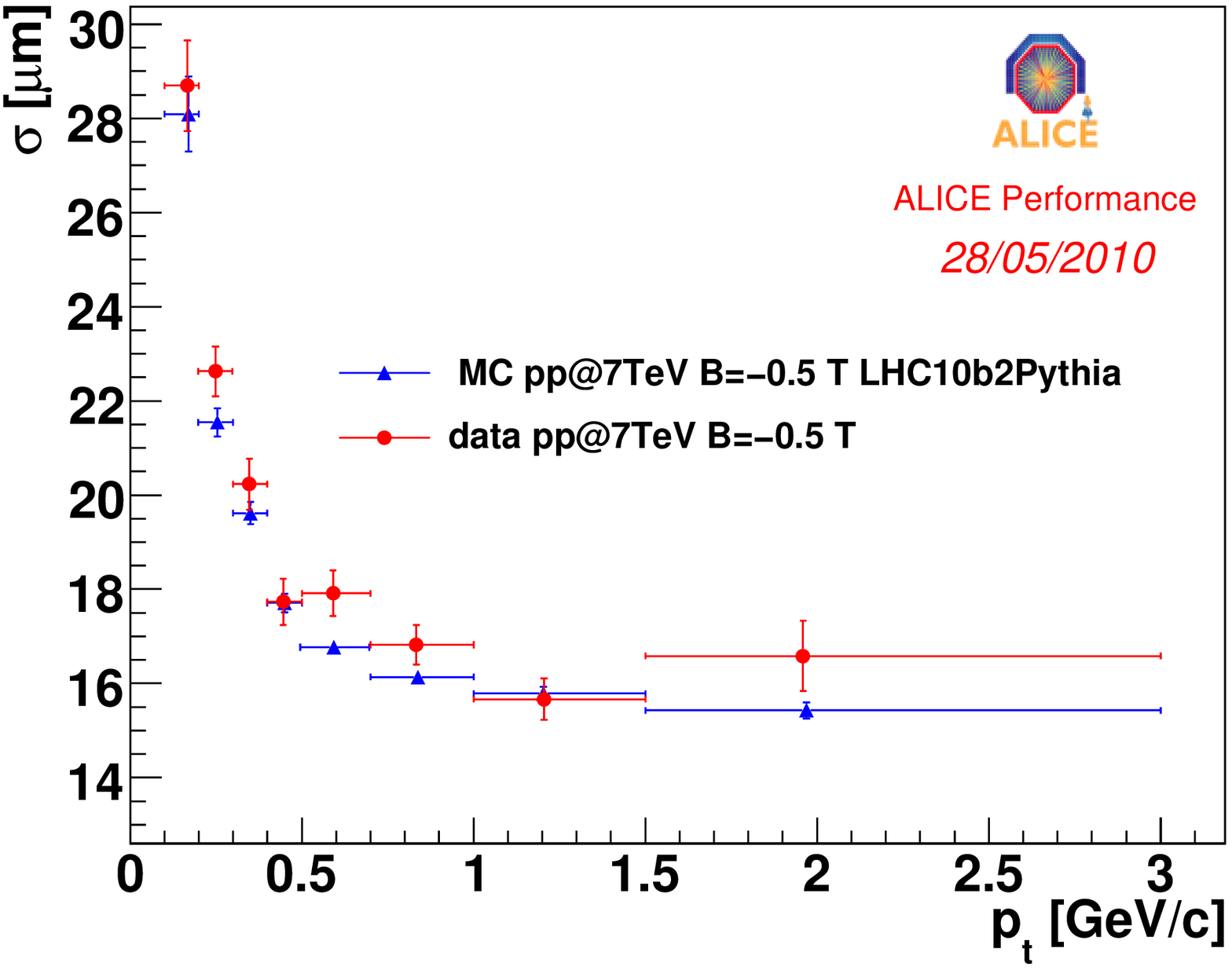}
      \caption[]{(left) spread of the SPD $\dxloc$ distribution as a function of the sum
	($\alpha_1+\alpha_2$) of the incidence angles on the two 
	overlapping modules, as obtained with tracks from cosmic-rays in 2008 (from~\cite{ITSalignArticle}).
	Right: spread of the $\dxloc$ as a function of $\pt$ for the inner SPD layer,
	obtained with data from \pp at $\sqrt{s}=7~\tev$.
      }
    \label{fig:AlignmentSPD}
  \end{center}  
\end{figure}
\subsubsection{Status of SPD alignment}
Two algorithms, both based on the 
minimization of track-to-point residuals, have been used to align the SPD
with tracks from cosmic-rays: the main
one, based on the Millepede algorithm~\cite{refMP1}, performs a global minimization
of all alignment and track parameters in the same time while the other performs a
module-by-module minimization and is iterated to account for the correlations 
among the parameters recovered for the different modules~\cite{ITSalignArticle,noteITSalign}.
The alignment quality
as well as the values of the alignment parameters obtained with the two methods
are comparable. 
About $80\%$ of the SPD modules were aligned within the target of 
containing the worsening of spatial resolution due to misalignment within $20\%$ of
the nominal resolution. 
The remaining modules have been aligned with \pp data.
Residuals from cosmic-ray tracks
and \pp tracks are used together, with different weights, in the Millepede
algorithm to exploit the different module correlations in the two samples. 
Figure~\ref{fig:AlignmentSPD} (left) shows the spread of the $\dxloc$
distribution as a function of the sum
($\alpha_1+\alpha_2$) of the incidence angles on the two 
overlapping modules, as obtained with tracks from cosmic-rays in 2008.
Monte Carlo simulation results are reported for comparison. 
The 2008 data are well
described by the simulation with a random residual misalignment with 
$\sigma\approx 7~\mum$.
On the right panel in the same figure, obtained with data from \pp at $\sqrt{s}=7~\tev$,
the spread of the $\dxloc$ is reported as a function of $\pt$. 
The rise at low $\pt$ is caused by multiple scattering, whose effect
becomes negligible for $\pt\gsim 1~\gev/c$. Data points are compatible within errors with 
the results from a Monte Carlo simulation 
obtained using random gaussian misalignments with $\sigma=7~\mum$. 
This value can be considered an estimate of the average level of residual
misalignment for all the SPD modules.
Further studies are now in progress to fully understand 
the intrinsic detector resolution and to address correlations
among misalignments, not detectable with this observable. 
\begin{figure}[!t]
  \begin{center}
    \vspace{-3mm}
    \includegraphics[height=.26\textheight]{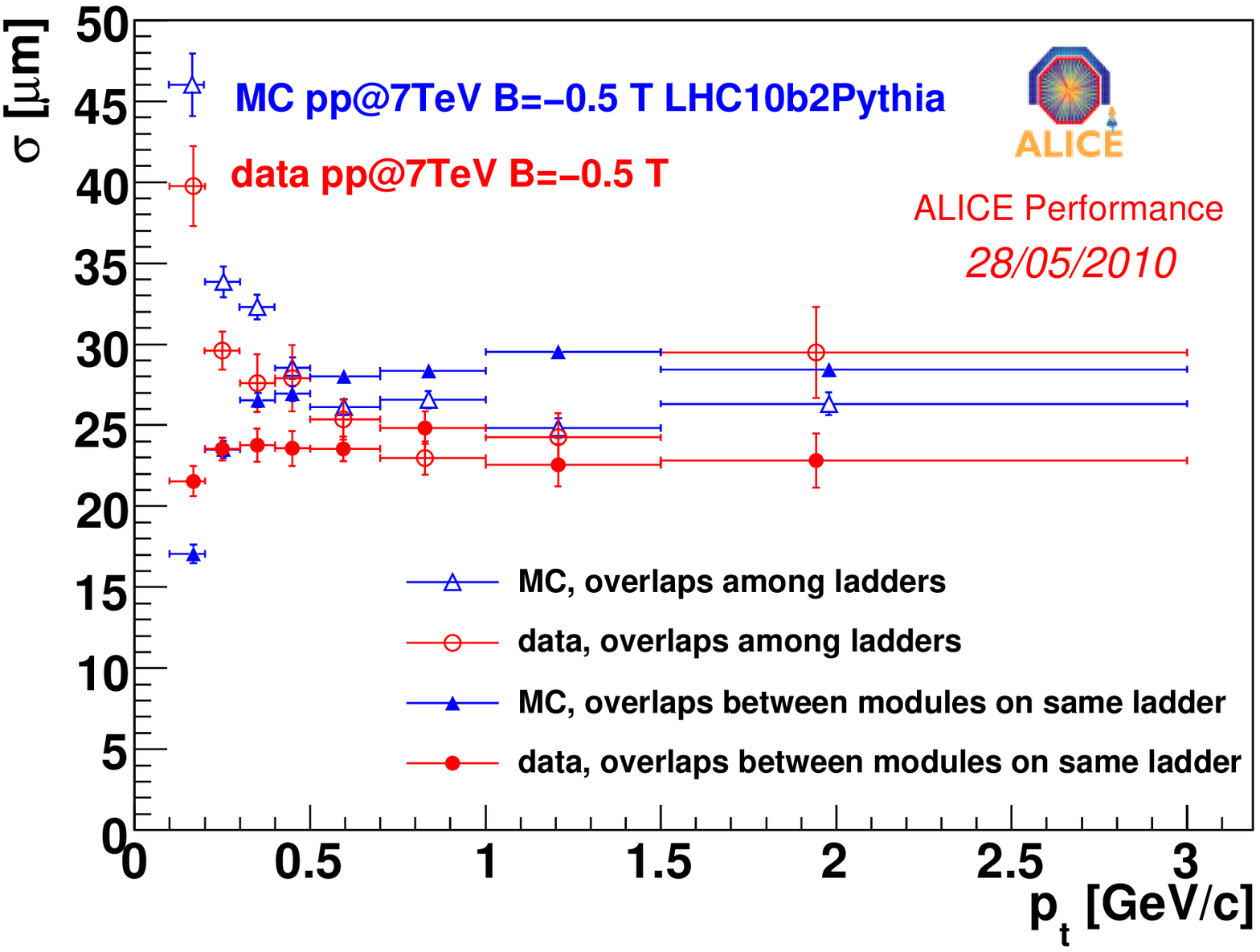}
    \includegraphics[height=.25\textheight]{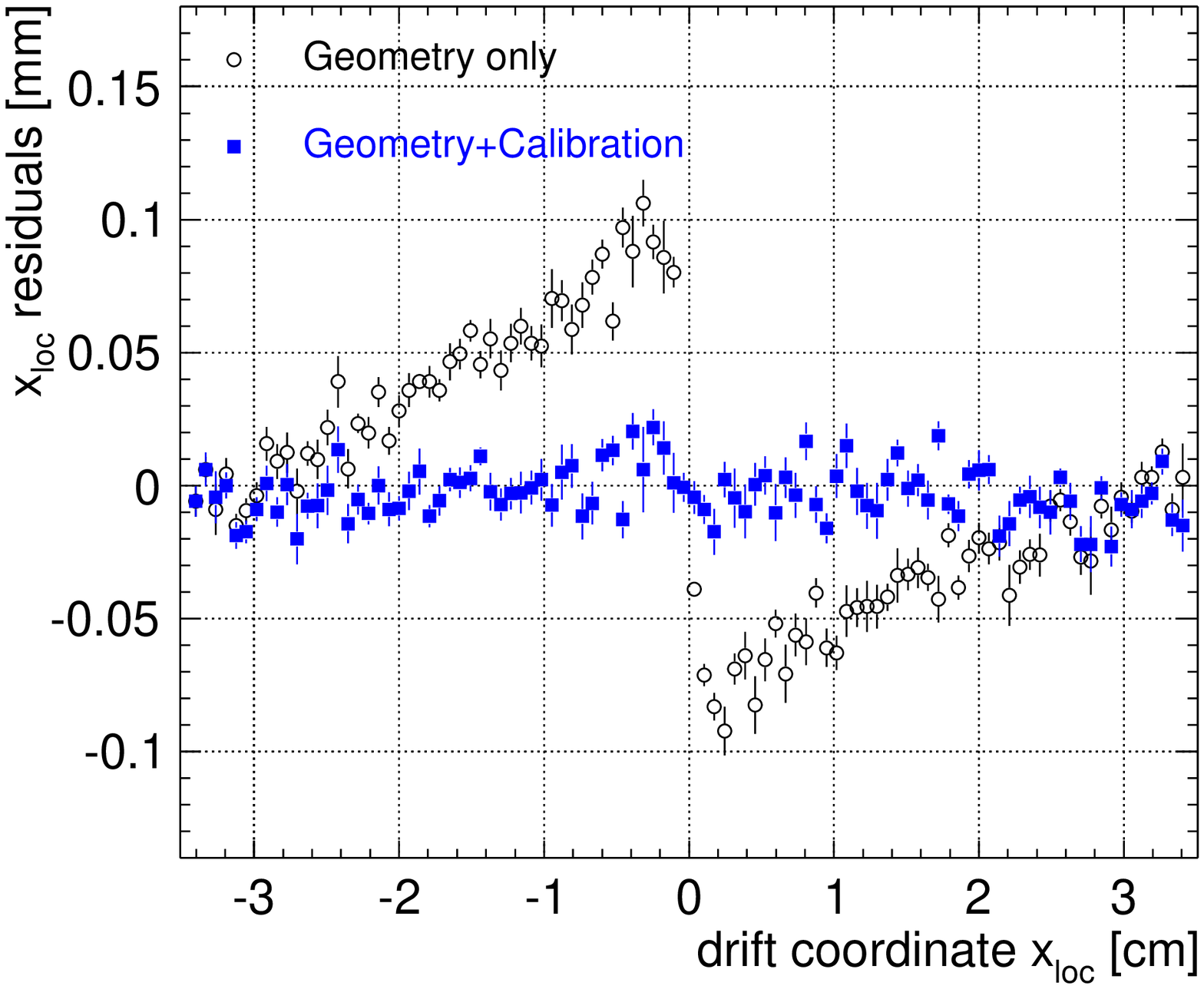}
    \caption[]{(left) spread of $\dxloc$ distribution obtained with \pp data at 7~$\tev$ 
      as a function of $\pt$ for overlaps between modules along the same ladder (filled markers) and  
      for overlaps between different ladders (empty markers).
      Right (from~\cite{ITSalignArticle}): residuals along the drift coordinate for one SDD module 
      as a function of drift coordinate after Millepede alignment
      with only geometrical parameters and with geometrical+calibration
      parameters.
    }
    \label{fig:AlignmentSSDSDD}
  \end{center}  
\end{figure}
\subsubsection{Status of SSD alignment}
The survey measurements performed during
SSD assembly and mounting phases assured a satisfactory first-alignment status for this detector.
The survey was carried out in two stages: the measurement of the
positions of the modules 
on the ladders and the measurement of the positions of the
ladder endpoints on the support cone. 
The modules are mounted with a small (2~mm) overlap in
both the longitudinal ($z$, modules on the same ladder) and transverse
directions ($r\varphi$, adjacent ladders). 
The residuals between
the ``extra'' points in these overlaps have 
been studied with tracks from cosmic-rays as well as from \pp
events. In Fig.~\ref{fig:AlignmentSSDSDD} (left)
the spread of $\dxloc$ distribution obtained with \pp data at 7~$\tev$ 
is shown as a function of $\pt$, separately
for overlaps between modules along the same ladder (filled markers) and  
for overlaps between different ladders (empty markers). For the latter, the effect of multiple scattering
at low $\pt$ is more evident because of the larger distance between the overlapping modules.
The distributions obtained with MC simulations seem to slightly overestimate 
the residual misalignment. The module and ladder residual misalignment 
in the simulations was set by extracting randomly,
according to a uniform distribution, smearing values for 
the module and ladder alignment parameters.
On the basis of the survey and of the alignment quality estimated
with cosmic-ray tracks, the values were chosen
so that the overall maximum displacement
could not exceed, for any point on a ladder,
$5~\mum$ in $x$, $10~\mum$ in $y$,
$5~\mum$ in $z$ due to module misalignment and
$10~\mum$ in $x$, $100~\mum$ in $y$,
$50~\mum$ in $z$ due to ladder misalignment.
We are now investigating whether the drop at low $\pt$, 
observed for overlaps on the same ladder, is related 
to a better determination of the cluster centre-of-gravity: 
due to the larger curvature, low $\pt$ tracks have a larger incidence angle 
than high $\pt$ tracks and can cross more neighbouring strips.
A study of track-to-point residuals and of
$\dxy$ distribution was also performed with 2008 
cosmic-ray data~\cite{SSDnote}.
The residual misalignment is estimated to be $\lsim 5~\mum$ for 
the module position on the ladder and $\lsim 10~\mum$ for the ladder positions.
\subsubsection{Status of SDD alignment}
The alignment of the SDD detectors
for the $x_{\rm loc}$ coordinate (reconstructed from the drift time)
is complicated by the interplay between the geometrical misalignment 
and the calibration of drift velocity and minimum drift time $t_0$. 
The $t_0$ parameter accounts for the delays between the time when 
a particle crosses
the detector and the time when the front-end chips receive the trigger 
signal.  The $t_0$ is determined by running the 
Millepede minimization with the $t_0$ as a free global 
parameter for each of the 260 SDD modules.
Similarly, the drift velocity is considered as a free parameter for those SDD modules
(about 35\%) with mal-functioning injectors~\cite{SittaThis}.
An example is shown for a specific SDD module in Fig.~\ref{fig:AlignmentSSDSDD} (right),
where the $x_{\rm loc}$ residuals along the drift direction
are shown as a function of $x_{\rm loc}$.
The clear systematic shift between the two drift regions ($x_{\rm loc}<0$
and $x_{\rm loc}>0$) is 
removed when also the calibration parameters are
fitted by Millepede (square markers). 
For the fraction of modules for which the drift velocity 
is well understood the current ``effective'' resolution on the local $x$ coordinate
has been evaluated to be $\sim 65~\mum$. We are now completing the alignment and calibration
of the remaining modules. 
\begin{figure}[!t]
  \begin{center}
    \includegraphics[height=0.23\textheight]{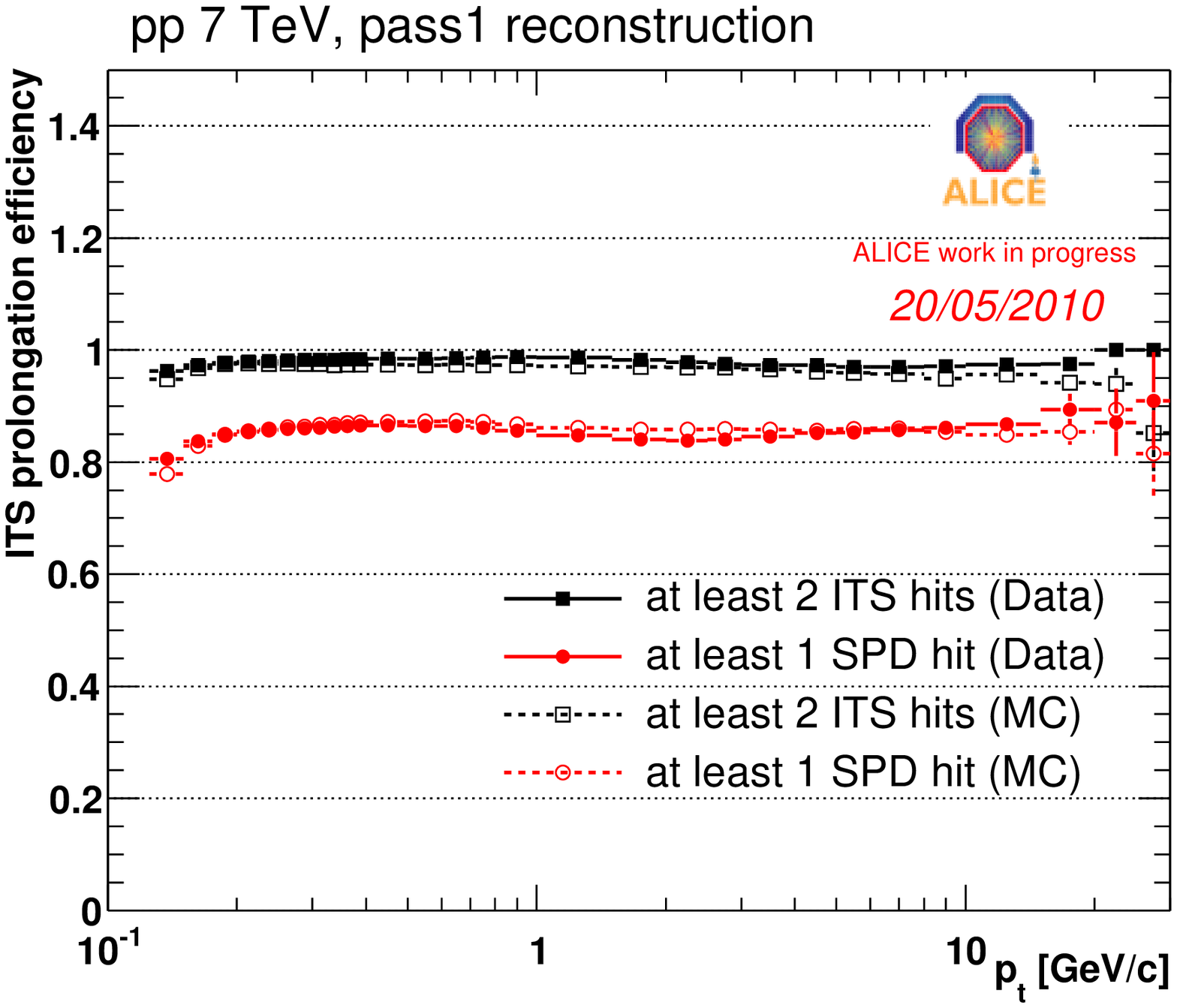}
    \includegraphics[height=0.23\textheight]{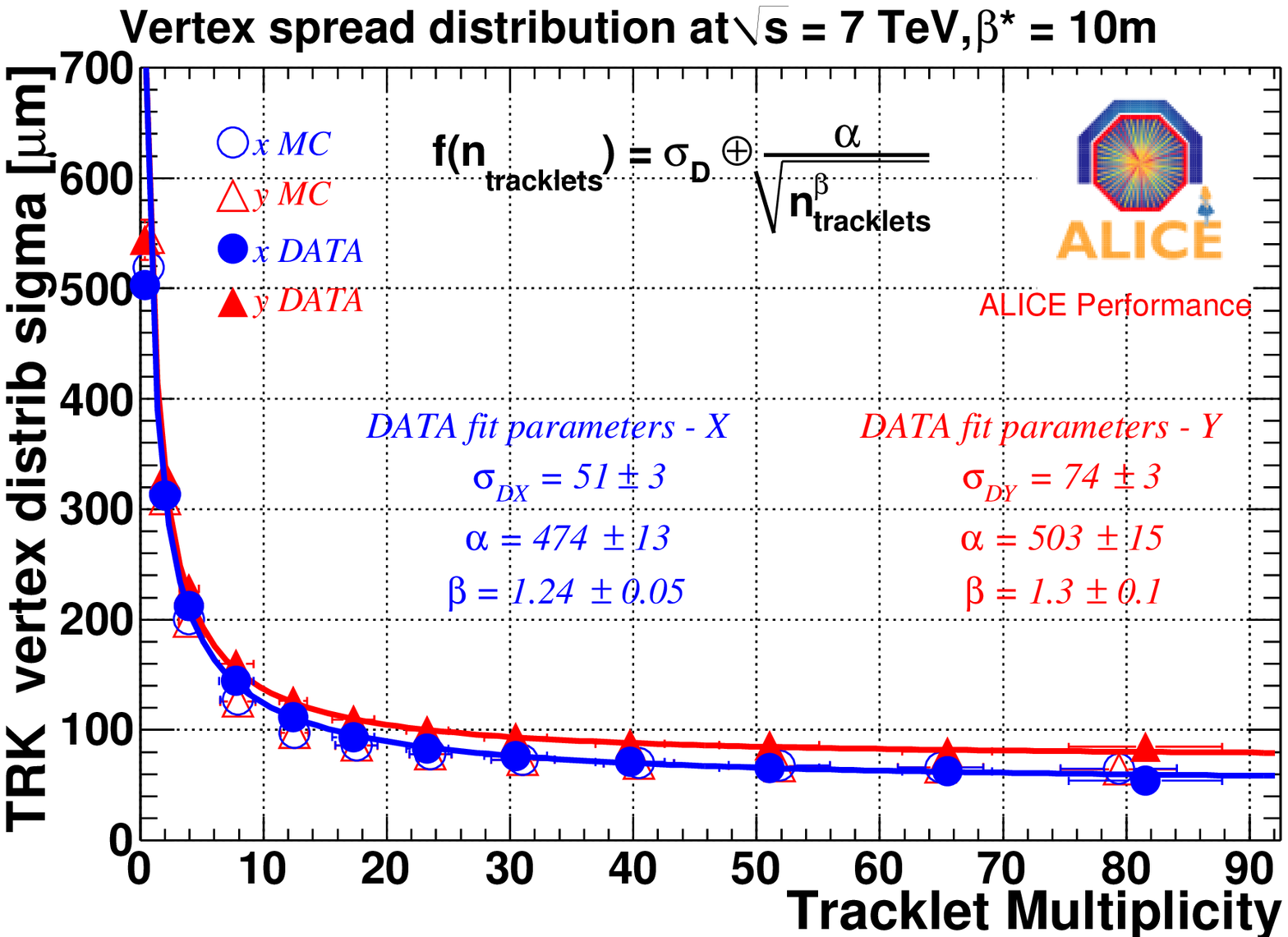}
    \caption{(left) prolongation efficiency between TPC and ITS as a function
      of transverse momentum in the case two points in the ITS are required (square markers) and with the further
      request that at least one point is in the SPD (round markers). Filled markers refers to data from
      \pp collisions at $7~\tev$ while empty markers refer to MC simulations. 
      Right: spread of $x$ and $y$ coordinates of the reconstructed primary vertex 
      for proton-proton collisions at $7~\tev$ for data (filled markers)
      and MC (empty markers). The asymptotic trend is determined
      by the size of the beam spot. The formula reported in the panel is used to
      fit data points (continuous lines).}
    \label{fig:Trk_Eff_VtxSpread}
    \end{center}
  \end{figure}
\subsection{Performance of track reconstruction}
Track reconstruction is performed via a Kalman filtering algorithm~\cite{aliceJINST,kalman_ALNOTE}. 
Charged tracks are parametrized by 5 parameters describing 
a helix. In the Kalman filtering approach the track trajectory is only
locally a helix: the track parameters are updated
at each point along the track trajectory allowing the possibility to
account for deviations due to the interaction with the material (energy loss,
multiple scattering). Starting from some initial approximation for the track parameters, 
points to be added to the track are looked for into a search-road determined by the uncertainty on the
track extrapolation at a given plane and on the cluster position uncertainty at the same plane.
An additional uncertainty on the ITS cluster position has been added to the intrinsic resolution
to account for residual misalignment. The adopted values,
reported in Table~\ref{tab:ResolMisalParamTracking},
were chosen in order to optimize track reconstruction efficiency and precision.
\begin{table}[!bt]
  \caption{Fraction of active modules (see~\cite{TurrisiThis,SittaThis}), average residual misalignment 
    and additional error added to the intrinsic cluster position uncertainty during track reconstruction
    to optimize tracking efficiency and precision.
  } 
  \begin{center}
    \begin{tabular}{cr|c|c|c|c|c|c}
                                             & &  SPD1 & SPD2 & SDD1 & SDD2 & SSD1 & SSD2 \\
      \hline
      \multicolumn{2}{c|}{active modules [\%]} & 75-80 &82-89 & 91& 90& 92& 89 \\
      \hline
      \multirow{2}*{residual misalignment level} & r$\phi$~[$\mum$] & <10 & <10 & 60 & 60 & <15 & <15 \\
                                          & z~[$\mum$] & negl. & negl. & 50 & 50 & ~100 & ~100 \\                                                                        
      \hline
      \multirow{2}*{additional error in tracking} & r$\phi$~[$\mum$] & 10 & 30 & 500 & 500 & 20 & 20 \\
                                                  & z~[$\mum$] & 100 & 100 & 100 & 100 & 500 & 500 \\                                                                         
      
      \end{tabular}
    \label{tab:ResolMisalParamTracking}
  \end{center}
\end{table}

Track reconstruction is performed in the following steps.
The first one is the computation of the primary vertex using SPD tracklets.
Track reconstruction then starts in the TPC (inward), 
using the
outermost pad rows and the computed primary vertex position as seed, 
and continues in the ITS (inward), matching TPC reconstructed tracks to the SSD layers and
following them down to the innermost SPD layer. 
The next step is track back-propagation, to the outermost layer of the ITS, to the outermost radius of the TPC and,
after the extrapolation and the track finding in the TRD, 
to the outer layers (TOF, HMPID, PHOS, EMCal) 
for Particle IDentification (PID). 
A ``refit'' is then performed: reconstructed tracks are re-fitted inward in TRD, TPC, ITS and are propagated to the primary
vertex reconstructed in the first step.
Finally the primary vertex is recalculated with optimal resolution using reconstructed tracks. 

To extend the $\pt$ acceptance down to $\sim 100~\mev/c$, 
an ITS-standalone tracking has been developed to reconstruct tracks in the ITS detector alone
~\cite{ITSstandalone}. 
In Fig.~\ref{fig:Trk_Eff_VtxSpread} (left) the prolongation efficiency between TPC and ITS is reported as a function
of transverse momentum. 
It is above $95\%$ and almost flat in the whole momentum range,
if tracks with at least two points in the ITS are considered. It drops down to
$\sim 85\%$ if the further request of one point in the SPD is done, due to the 
inactive modules in this detector (Table~\ref{tab:ResolMisalParamTracking}). MC simulations reproduce the efficiency
trend observed in data within $\lesssim 2\%$ in the range $0.2<\pt<10~\gev/c$.
\begin{figure}[!t]
  \begin{center}
\vspace{-2mm}
  \includegraphics[width=0.45\textwidth]{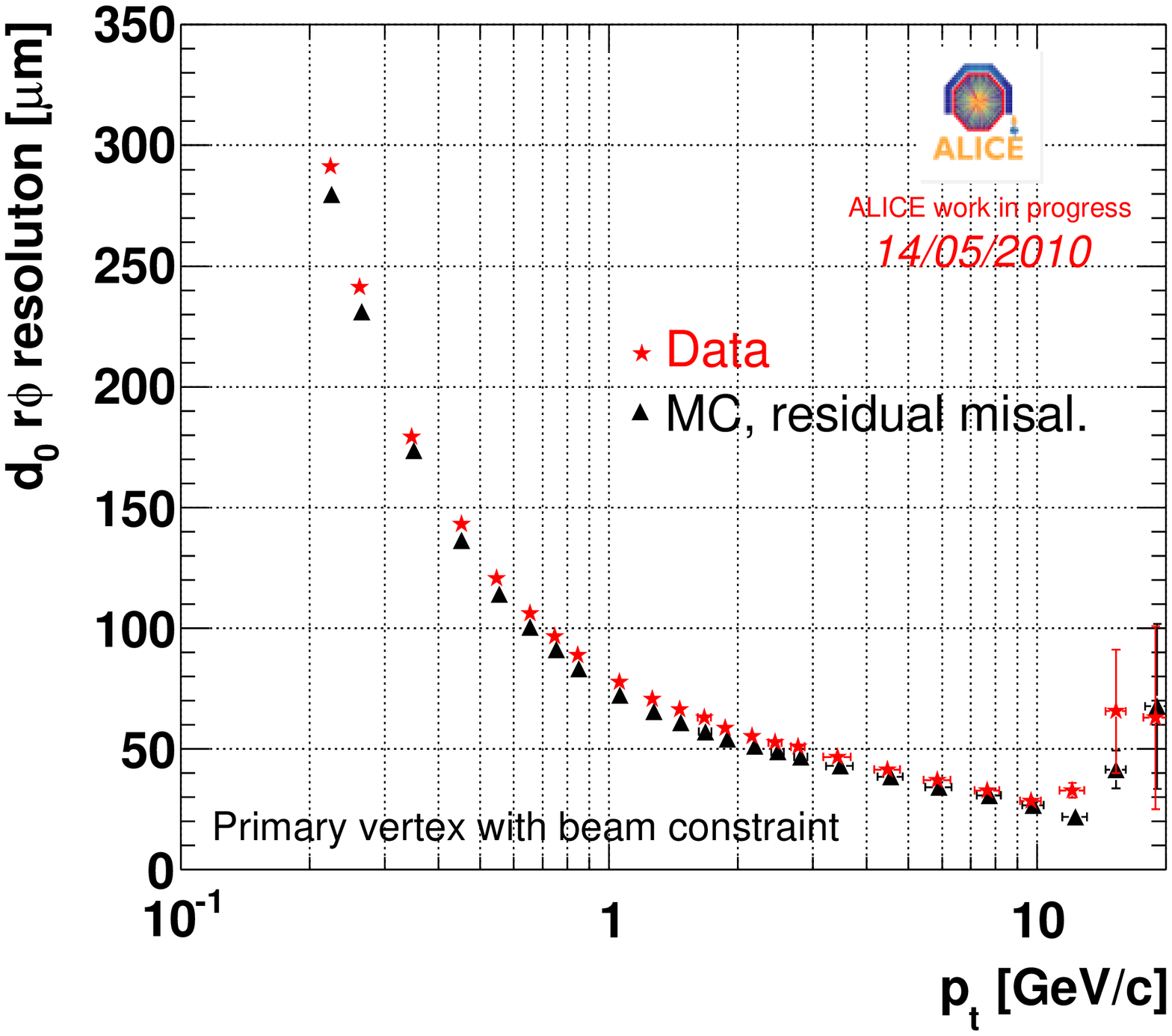}
  \includegraphics[height=0.245\textheight]{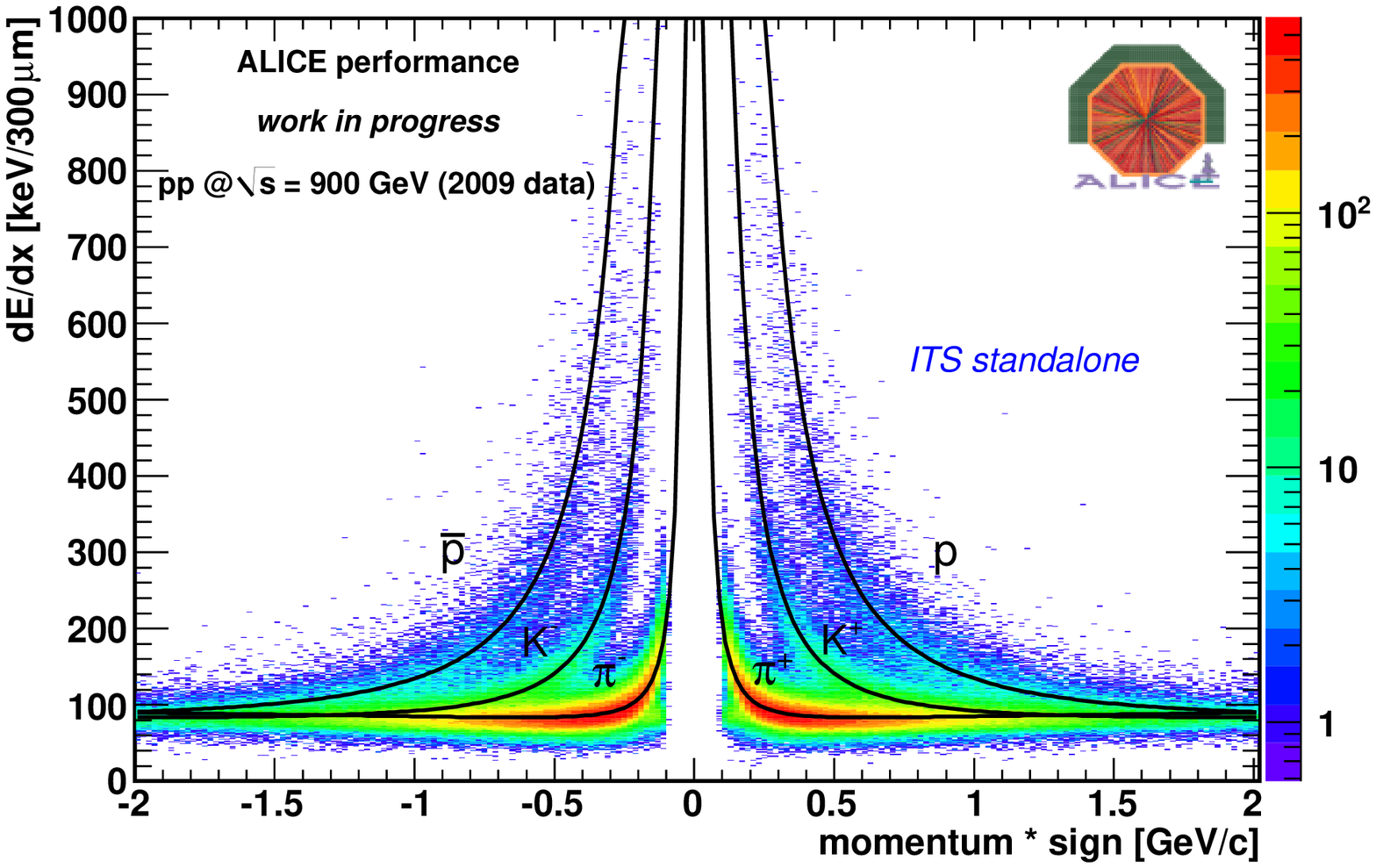}
  \caption{(left) transverse impact parameter resolution estimate. 
    Tracks satisfying the standard TPC track 
    quality cuts and with 2 points in the SPD were considered.
    For each track, the impact parameter was estimated with respect to 
    the primary vertex reconstructed without using the same track. 
    The primary vertex reconstruction does use the beam constraint.
    Right: truncated mean $\dedx$ as a function of the
    particle momentum times the charge. Continuous lines represent the theoretical
    average energy loss for kaon, pions and proton as calculated
    with the Bethe-Bloch formula.}
  \label{fig:d0resolpt}
  \end{center}
\end{figure}
\subsection{Performance of primary vertex reconstruction}
The z coordinate (along the beam line) 
of the interaction point is distributed in a range of several centimetres.
The beam spot size in the transverse plane varies between ten and hundred microns depending
on the beam optics ($\beta^{\star}$) and energy.  
Two algorithms~\cite{IntNotVtx_BrDaPrMa} are used to reconstruct the primary vertex position.
The first exploits the correlation between pairs of points
in the SPD ("tracklets"). The efficiency reaches $\sim 90\%$ with a tracklet
multiplicity of 4 and approaches 100\% when the tracklet multiplicity is 8. 
When the calculation of the full 3D position fails, the z coordinate alone
is calculated: in this case one tracklet is sufficient to get a 100\% efficiency.

The second algorithm is based on the straight line approximation of fully
reconstructed tracks in the vicinity of the vertex. 
With a tracklet multiplicity of 3 
the reconstruction efficiency is $\sim 95\%$ if the information of the beam spot position
and size is used to constrain the vertex, $\sim 80\%$ otherwise; 100\% is 
approached at higher multiplicity. 
In Fig.~\ref{fig:Trk_Eff_VtxSpread} (right)   
the spread of the reconstructed {\it x} and {\it y} coordinates  
as a function of the SPD tracklet multiplicity in the event is reported. 
The curves flatten
with higher multiplicity, corresponding to a better determination of the 
primary vertex position. Monte Carlo points are in good agreement with data.
Data points are fitted with the function
$\sigma_{D}\oplus \frac{\alpha}{\sqrt{N_{trkl}^{\beta}}}$:
$\sigma_{D}$ describes the asymptotic value of the curve,
determined by the size of the beam interaction diamond along
the considered axis, while the second term
represents a 
parametrisation of the 
primary vertex resolution as a function of the tracklet multiplicity.
\subsection{Impact parameter resolution}
The distribution of the track impact parameter with respect 
to the reconstructed primary vertex is the convolution 
of a detector resolution function with the unknown true impact parameter distribution,  
the latter giving rise to relevant tails due to secondary particles. 
Therefore, in order to estimate the resolution on the impact parameter,
only primary tracks should be considered. 
It has been checked that the contribution of secondaries 
in a range $\approx 2$~RMS produces almost negligible effects 
on the standard deviation obtained by fitting the impact parameter
distribution in this range. 
This standard deviation can be considered a good estimate of the impact parameter resolution.
Recently we considered the opportunity to fit the distribution in a larger range
using an exponential function to describe the tails. The two methods give compatible results.
To obtain an unbiased evaluation of the impact parameter, 
the primary vertex position is recalculated track-by-track
excluding the current track from the computation. 
In Fig.~\ref{fig:d0resolpt} (left) the impact parameter resolution as a function of
transverse momentum is shown. The obtained curve is the result of the convolution 
of the track position and the primary vertex resolutions. 
The agreement with MC is within $10\%$.
Further improvement is expected with alignment fine-tuning 
and when the material budget will be under control to per cent level.
\section{Particle Identification with the SSD and SDD}
\label{sec:ITSpid}
The analogue readout of the electronic signal produced 
in the SSD and SDD detector allows for the
measurement of the energy loss ($\dedx$) in the silicon layers~\cite{SittaThis}. This information
is used for particle identification at low momentum, in the 
non-relativistic region. 
Since the energy losses are distributed
according to the Landau distribution, characterized by a long tail towards
high energy loss values, a truncated mean $\dedx$ is performed. 
In Fig.~\ref{fig:d0resolpt} (right) the 
truncated mean $\dedx$ is shown as a function of momentum times the particle charge. 
A special correction for the different path
lengths inside the silicon has been applied. 
The bands of kaons, pions and protons are clearly
visible and centred around the continuous lines representing the
theoretical curves from the Bethe-Block formula for the average energy loss.
The ITS allows for hadron separation below $100~\mev/c$, a region outside
of TPC and TOF PID capability. Good separation between pions and kaons is achievable up to $\sim 0.5~\gev/c$
and between proton and pions up to $\sim 1~\gev/c$.  
\begin{figure}[!t]
  \begin{center}
    \includegraphics[height=.23\textheight]{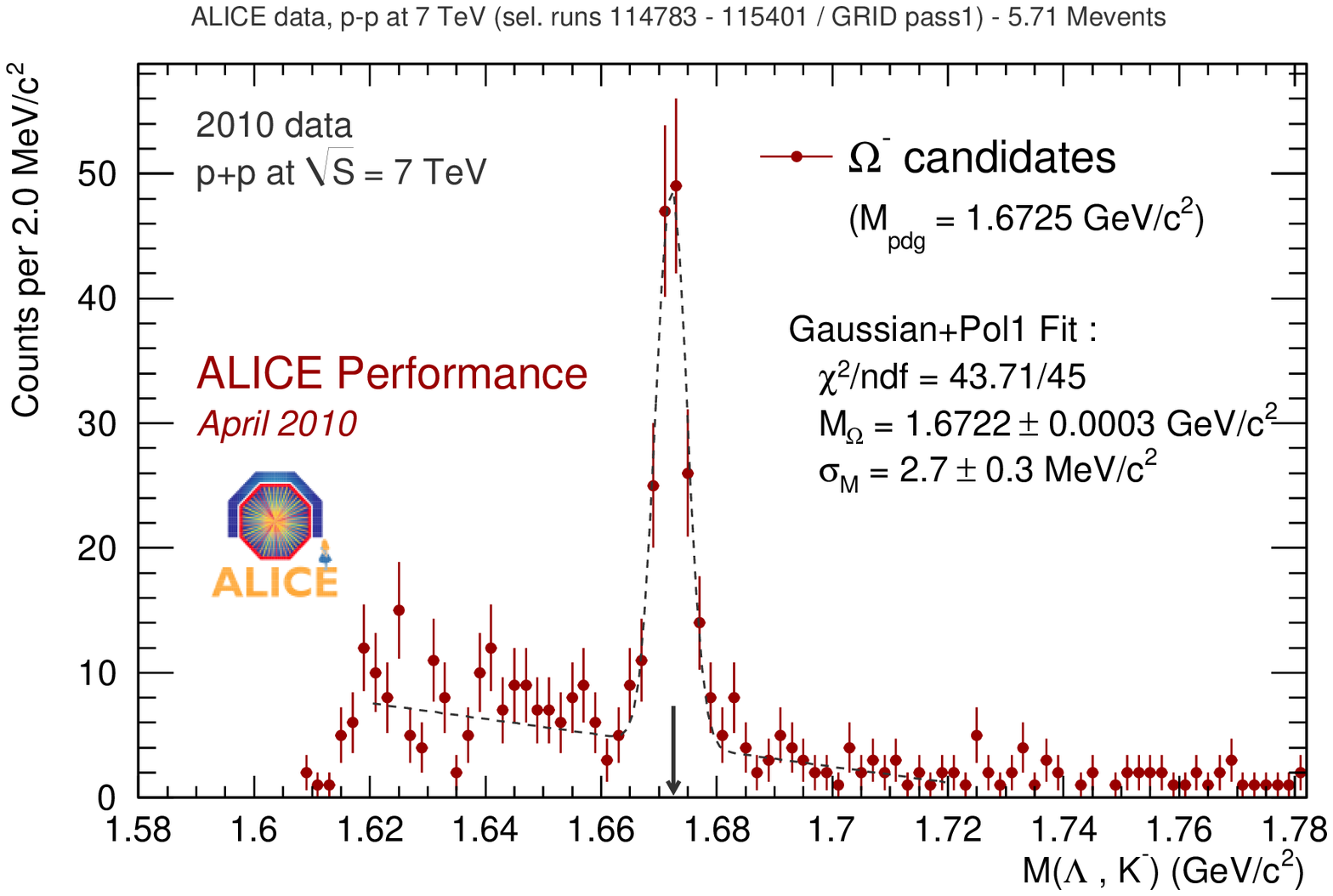}
    \includegraphics[height=.23\textheight]{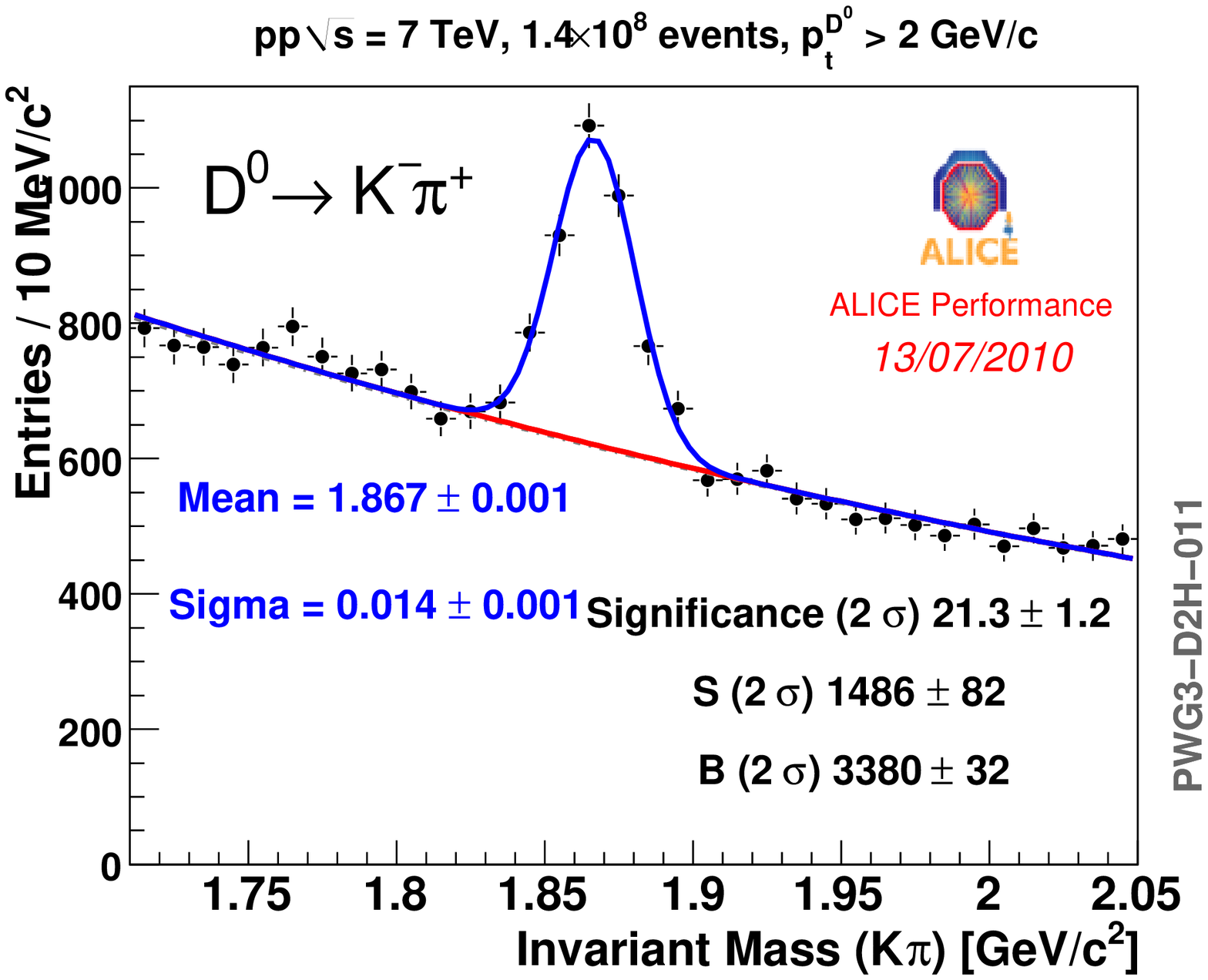}
    \caption[]
	    {(left) $\Omega$ candidates invariant mass distribution 
	      obtained with $\sim 5.7\cdot 10^{6}$ minimum bias events from \pp collisions at
	      $7~\tev$. Right: invariant mass distribution 
	      of $\dokpi$ candidates for $\pt^{D^0}> 2~\gev/c$ obtained 
	      with $\sim 1.4\cdot 10^{8}$ minimum bias events. 
	    }
	    \label{fig:Performance}
  \end{center}  
\end{figure}
\section{Examples of physics performance: rediscovery of the familiar strange and charm world}
\label{sec:physperformance}
Since the first \pp collisions at $900~\gev$ in November 2009, the ITS has covered
an important role in all the physics analyses ongoing, from the study
of the global properties of the event (multiplicity, $\pt$ spectra, proton/anti-proton ratio)
to the reconstruction of V0 topologies and cascades up to the
``rediscovering'' of charmed mesons. In Fig.~\ref{fig:Performance} (left)
an example of the invariant mass distribution of $\Omega$ candidates 
obtained with $\sim 5.7\cdot 10^{6}$ minimum bias events from \pp collisions at
$7~\tev$ is shown. 

On the right panel in the same figure, the invariant mass distribution
of $\dokpi$ candidates for $\pt^{D^0}> 2~\gev/c$ is shown as obtained 
with $\sim 1.4\cdot 10^{8}$ minimum bias events. 
The $\dokpi$ (branching ratio $\approx 3.8\%$) is among the most promising 
channels for the measurement of open charm production at mid-rapidity. The detection
strategy is based on an invariant mass analysis of all the pairs of tracks
with opposite charges. To cope with the large combinatorial background, 
displaced-vertex topologies are looked for, i.e. tracks displaced 
from the primary vertex are selected and good alignment between the reconstructed
meson momentum and its flight direction is required. 
At the moment of writing this proceeding ALICE observed
the signal of charm mesons in the  
$\dpluskpipi$, $\dstardopi$ ($\dokpi$), $\dokpipipi$ decay channels,
in the range $1 < \pt < 12~\gev/c$.
Precise reconstruction of the primary vertex and of the track trajectory
in the vicinity of the vertex are essential prerequisites to perform these 
analyses: the performance of the ITS is thus crucial.

\section{Summary}
After a two year long phase of commissioning, started with 2008 cosmic-ray run
and finalized with \pp collisions, the ITS is in a good shape as
far as concern the understanding of the
detector response, the alignment and the description of the
material budget. The performance on the track and vertex reconstruction
are close to the nominal one and reproduced by MC simulation
within a few percent. The ITS capability to identify particles down
to $\sim 0.1~\gev/c$ has been shown. 
The ITS will cover a fundamental role in the fulfilment of 
ALICE physics targets as already done for the first three ALICE
papers~\cite{ALICEmultart} based entirely on the ITS and trigger detectors.

\end{document}